\documentclass[twocolumn,secnumarabic,amssymb,amsmath, nobibnotes, aps, prd]{revtex4}
\usepackage{graphicx}
\usepackage{dcolumn}
\usepackage{bm}

\begin{document}

\title{Chirped-pulse oscillators: a unified standpoint}

\author{V. L. Kalashnikov}
\affiliation{Institut f\"{u}r Photonik, TU Wien, Gusshausstr.
27/387, A-1040 Vienna, Austria} 
\author{A. Apolonski}
\affiliation{Department f\"{u}r Physik der
Ludwig-Maximilians-Universit\"{a}t M\"{u}nchen, Am Coulombwall 1,
85748, Germany} \affiliation{Institute of Automation and
Electrometry, RAS, 630090 Novosibirsk, Russia}

\begin{abstract}
A completely analytical and unified approach to the theory of
chirped-pulse oscillators is presented. The approach developed is
based on the approximate integration of the generalized nonlinear
complex Ginzburg-Landau equation and demonstrates that a
chirped-pulse oscillator is controlled by only two parameters. It
makes it easy to trace spread of the real-world characteristics of
both solid-state and fiber oscillators operating in the positive
dispersion regime.
\end{abstract}


\maketitle

\section{\label{intro}Introduction}

High-energy laser oscillators nowadays allow high-intensity
experiments such  as direct gas ionization \cite{morgner}, where the
level of intensity must be of the order of 10$^{14}$ W/cm$^2$.  One
can expect soon pump-probe diffraction experiments with electrons
and direct high-harmonic generation in gases and production of
nm-scale structures at the surface of transparent materials. Each
case calls for an intensity of the order of that demonstrated above
or higher, which means generating tens  up to hundreds of $\mu$J
pulses at the fundamental MHz repetition rate of an oscillator
\cite{keller1}. From the examples above it  can be seen that high
repetition rates are preferable to kHz rates (now commercially
available) because the signal rates in, for example, electron
experiments are usually low and an improvement factor of 10$^3$ --
10$^4$ due to the higher repetition rate of the pulses significantly
enhances the signal-to-noise ratio. In addition to this physical
factor, existing kHz systems are more expensive, complex and less
stable.

There are a few ways of increasing the oscillator pulse energy $E$,
which is a product of the average power and the repetition period:
by increasing the cavity length and/or increasing the power
\cite{fujimoto,krausz2,keller2,fernandez2,keller3,neuhaus}. The
catch is that a long-cavity oscillator suffers from instabilities
owing to nonlinear effects caused by the high pulse peak power
$P_0$. The leverage is to stretch a pulse and thereby decrease its
peak power below the instability threshold. Recent progress
demonstrating the feasibility of this approach has been achieved for
Ti:sapphire oscillators operating in both the negative- (NDR)
\cite{fujimoto2,fujimoto3} and positive-dispersion regimes (PDR)
\cite{fernandez,naumov},  for near-infrared Yb-doped solid-state
oscillators operating in both the NDR
\cite{keller2,neuhaus,morgner2} and the PDR
\cite{morgner2,morgner3}, and for fiber oscillators operating in the
all-normal dispersion (ANDi) (i.e. positive dispersion) regime
\cite{wise3,limpert}.

The fundamental difference between the NDR and PDR is that, in
the former, the Schr\"{o}dinger soliton develops \cite{kartner1}.
The soliton width $T$ and energy $E$ can be expressed as
\cite{agrawal}

\begin{equation} \label{soliton}
T = \sqrt {{{\left| \beta  \right|} \mathord{\left/
 {\vphantom {{\left| \beta  \right|} {\gamma P_0 }}} \right.
 \kern-\nulldelimiterspace} {\gamma P_0 }}} ,\,\,E = {{2\left| \beta  \right|} \mathord{\left/
 {\vphantom {{2\left| \beta  \right|} {\gamma T,}}} \right.
 \kern-\nulldelimiterspace} {\gamma T,}}
\end{equation}

\noindent where $\gamma$ is the self-phase modulation (SPM)
coefficient of a nonlinear medium (active crystal, fiber, air,
etc.), and $\beta$ is the net-group-delay-dispersion (GDD)
coefficient of an oscillator cavity. Since the peak power $P_0$ has
to be kept lower than the threshold value $P_{th}$ in order to avoid
soliton destabilization, one can estimate the maximum attainable
energy as $E = 2 P_{th} T$. Energy scaling thus requires pulse
stretching. However, the latter results from the substantial GDD
growth [quadratically with energy, see Eq. (\ref{soliton})]:

\begin{equation} \label{soliton2}
\left| \beta  \right| = {{E^2 \gamma } \mathord{\left/
 {\vphantom {{E^2 \gamma } {4P_{th} .}}} \right.
 \kern-\nulldelimiterspace} {4P_{th} .}}
\end{equation}

\noindent Hence, the pulse width increases linearly with energy
(correspondingly, the spectrum narrows with $E$). As a result, i)
energy scaling requires a huge negative GDD, ii) the soliton obtained
has a large width, and iii) it is not compressible linearly.

In contrast to the soliton regime, the pulse is stretched in the PDR
\cite{proctor} and its peak power is reduced due to chirp $\psi$
\cite{haus2,kalash1}. The chirp compensates the narrowing of the spectrum
with energy \cite{kalash1}:

\begin{equation} \label{width}
2\Delta  \approx {{8\pi \psi } \mathord{\left/
 {\vphantom {{8\pi \psi } {\gamma E,}}} \right.
 \kern-\nulldelimiterspace} {\gamma E,}}
\end{equation}

\noindent where $\Delta$ is the spectrum half-width. The advantage
is that such a pulse [chirped solitary pulse (CSP)] is compressible
linearly down to $T \approx 2/\Delta$ (the compression factor is
$\approx\psi$).

As was found, the CSP can be described as a solitary pulse solution
of the cubic nonlinear complex Ginzburg-Landau equation (CGLE)
\cite{haus2} or, more generally, the cubic-quintic nonlinear CGLE
\cite{kalash1,akh1,kalash2,wise1}. This equation is the generalized
form of the master mode-locking equation
\cite{kartner1,haus2,akhmed,haus3}, which provides an adequate
description of mode-locked oscillators (both fiber and solid-state
ones). Furthermore, the nonlinear CGLE is used in quantum optics,
modeling of Bose-Einstein condensation, condensed-matter physics,
study of non-equilibrium phenomena, and nonlinear dynamics, quantum
mechanics of self-organizing dissipative systems, and quantum field
theory \cite{kramer}. Therefore, analysis of the CSP solutions of
nonlinear CGLE is of interest not only from the practical but also
from the theoretical point of view.

Since the underlying problem is multiparameter and not integrable in
the general form, there is no a uniform standpoint on  CSPs
developing in, for example, the ANDi fiber oscillator \cite{wise2}
and the chirped-pulse oscillator (CPO) \cite{naumov}. In particular,
the physical parameters of oscillators vary greatly and it is not
clear \emph{a priori} whether the mechanisms governing the PDR are
unified.

In this work, we propose an  approximate method of integrating the
generalized nonlinear CGLE and show that the CSP is its
two-parametrical solitary pulse solution. As a result, the CSP
characteristics are easy to trace on a two-dimensional diagram
(``master diagram''). Comparison of the PDR parameters demonstrates
that the CSPs formed in the ANDi fiber oscillator and in the CPO i)
lie within distinct sectors of the unified master diagram, ii)
belong to mainly distinct branches of the solution, and iii) vary
with the parameters in different ways. The variation of the main CPO
characteristics (spectrum shape and width as well as pulse
stability) with the PDR parameters is analyzed along with the
numerical and experimental results. A comparison of the models based
on the different versions of the master equation is made.

\section{\label{s1}CSP solution of the generalized nonlinear CGLE}

The  evolution of the visible, near- and mid-infrared
electromagnetic fields in an oscillator can be described on the basis of the
slowly-varying field approximation, when the spectral width is much
smaller than the carrier frequency of the field. The field envelope $A$
is affected mainly by i) GDD, ii) SPM (non-dissipative factors), as
well as iii) saturable gain and linear loss, iv) spectral filtering,
and v) self-amplitude modulation (SAM) (dissipative factors)
\cite{kartner1,akhmed}. When the effects of higher-order dispersion
\cite{oe} and the field variation along a single oscillator round-trip
\cite{kalash1} are negligible, the oscillator dynamics obeys the
nonlinear CGLE:

\begin{equation}\label{GL}
\frac{{\partial A}} {{\partial z}} = \sigma A + \left( {\alpha  +
i\beta } \right)\frac{{\partial ^2 A}} {{\partial t^2 }} + \left(
{\frac{{\mu \varsigma }} {{1 + \varsigma \left| A \right|^2 }} -
i\gamma } \right)\left| A \right|^2 A,
\end{equation}

\noindent where $z$ is the propagation distance normalized to the
oscillator length (for a ring oscillator model; or to double its
length for a linear oscillator), $t$ is the local time [the
reference frame is co-moving with a solitary pulse solution of Eq.
(\ref{GL})], $\sigma = g - l - \mu$ is the dimensionless saturated
net gain ($g$, $l$ and $\mu$ are the saturated gain, non-saturable
and saturable loss coefficients, respectively), $\alpha$ is the
squared inverse transmission bandwidth of an oscillator, $\beta$ is
the GDD coefficient, $\varsigma$ is the inverse saturation power of
the self-amplitude modulator, $\gamma$ is the net SPM coefficient.
The normalization of the slowly varying field amplitude $A(z,t)$ is
chosen such that $P(z,t) \equiv \left| A \right|^2$ is the
instantaneous power. The SAM corresponds to a perfectly saturable
absorber. Such a character of SAM corresponds to a semiconductor
saturable mirror (SESAM). SESAMs are extensively used in the sub-
and over-$\mu$J oscillators operating in both PDR \cite{naumov} and
NDR \cite{neuhaus}. Such a technique provides stability and
self-staring ability of mode-locking and, to date, has no
alternative for high-energy femtosecond lasers. The main condition,
which allows describing the SESAM response in the form of
(\ref{GL}), is excess of the pulse width (few picoseconds for the
PDR) over the SESAM relaxation time ($\approx$100 fs) \cite{silb}.

Rescaling $z'=z \mu$, $t'=t \sqrt {\mu /\beta }$, and $P'=\varsigma
P$ demonstrates that Eq. (\ref{GL}) is in fact three-parametrical.
However, as  will be demonstrated below, the CSP is the
two-parametrical solution of Eq. (\ref{GL}). In contrast to Ref.
\cite{wise1}, we do not impose restrictions on the pulse phase
$\phi(t)$:

\begin{equation}\label{sol}
A\left( {z,t} \right) = \sqrt {P\left( t \right)} \exp \left( {i\phi
\left( t \right) - iqz} \right),
\end{equation}

\noindent where  $q$ is the phase produced by slip of the carrier
phase in relation to  the slowly varying envelope $\sqrt{P\left( t
\right)}$ \cite{kartner1}.

By analogy with Refs. \cite{kalash1,kalash2}, substitution of Eq.
(\ref{sol}) in Eq. (\ref{GL}) with the subsequent assumptions
$\alpha \ll \beta$ (GDD prevails over spectral dissipation), and $
{{d^2 \sqrt P } \mathord{\left/
 {\vphantom {{d^2 \sqrt P } {dt^2  \ll 1}}} \right.
 \kern-\nulldelimiterspace} {dt^2  \ll 1}}
$ (that is,  the adiabatic approximation $\beta \ll T^2$) results in
\begin{eqnarray}\label{sol1}
\begin{gathered}
  \gamma P\left( t \right) = q - \beta \Omega \left( t \right)^2 , \hfill \\
\beta \frac{{d\Omega \left( t \right)}} {{dt}} + \beta \frac{{\Omega
\left( t \right)}} {{P\left( t \right)}}\frac{{dP\left( t \right)}}
{{dt}} = \sigma  + \frac{{\mu \varsigma P\left( t \right)}} {{1 +
\varsigma P\left( t \right)}} - \alpha \Omega \left( t \right)^2, \hfill \\
\end{gathered}
\end{eqnarray}

\noindent where $\Omega \left( t \right) \equiv {{d\phi \left( t
\right)} \mathord{\left/
 {\vphantom {{d\phi \left( t \right)} {dt}}} \right.
 \kern-\nulldelimiterspace} {dt}}$ is the instant frequency.

Eq. (\ref{sol1}) results in

\begin{widetext}
\begin{equation}
\beta \frac{{d\Omega }} {{dt}} = \frac{{\left( {\Delta ^2  - \Omega
^2 } \right)\left[ {\left( {\Delta ^2  - \Omega ^2 } \right)\left(
{\mu  + \sigma } \right) - \alpha \Omega ^2 \left( {\Delta ^2  -
\Omega ^2  + {\gamma \mathord{\left/
 {\vphantom {\gamma  {\varsigma \beta }}} \right.
 \kern-\nulldelimiterspace} {\varsigma \beta }}} \right) + {{\sigma \gamma } \mathord{\left/
 {\vphantom {{\sigma \gamma } {\varsigma \beta }}} \right.
 \kern-\nulldelimiterspace} {\varsigma \beta }}} \right]}}
{{\left( {\Delta ^2  - \Omega ^2  + {\gamma \mathord{\left/
 {\vphantom {\gamma  {\varsigma \beta }}} \right.
 \kern-\nulldelimiterspace} {\varsigma \beta }}} \right)\left( {\Delta ^2  - 3\Omega ^2 } \right)}}. \label{sol1bis}
\end{equation}
\end{widetext}

The regularity condition $ {{d\Omega } \mathord{\left/
 {\vphantom {{d\Omega } {dt < \infty }}} \right.
 \kern-\nulldelimiterspace} {dt < \infty }}$ requires that the expression in square
 brackets in Eq. (\ref{sol1bis}) equals to zero when
 $\Omega(t)^2=\Delta^2/3$. This implies the expression for
 $\Delta^2$. Simultaneously, the condition $P(t)>$0
 requires the spectrum truncation so that $\Omega(t)^2<\Delta^2\equiv
 q/\beta$.  Thus, one arrives at the equations

\begin{eqnarray}\label{sol2}
\begin{gathered}
\varsigma P\left( 0 \right) = \frac{{\alpha \Delta ^2 }} {{\mu b}} =
\frac{3} {{4b}}\left[ {2\left( {1 + a} \right) - b \pm \sqrt
\Upsilon  } \right], \hfill \\
  \frac{{d\Omega }}
{{dt}} = \frac{\alpha } {{3\beta }}\frac{{\left( {\Xi ^2  - \Omega
^2 } \right)\left( {\Delta ^2  - \Omega ^2 } \right)}} {{\Delta ^2
- \Omega ^2  + {\gamma  \mathord{\left/
 {\vphantom {\gamma  {\varsigma \beta }}} \right.
 \kern-\nulldelimiterspace} {\varsigma \beta }}}}, \hfill \\
\frac{{\alpha \Xi ^2 }} {\mu } = \frac{{2\alpha }} {{3\mu }}\Delta
^2 + 1 + a + b.\hfill \\
\end{gathered}
\end{eqnarray}

\noindent Here $a\equiv \sigma/\mu$, $b\equiv \alpha \gamma/\beta
\varsigma \mu$, $ \Upsilon \equiv \left( {b - 2} \right)^2  +
4a\left( {2 + a + b} \right) $. The second equation in (\ref{sol2})
results from regularization of Eq. (\ref{sol1bis}) under condition
of $ {{d\Omega } \mathord{\left/
 {\vphantom {{d\Omega } {dt < \infty }}} \right.
 \kern-\nulldelimiterspace} {dt < \infty }}$.

Positivity of $\Upsilon$ requires that
 $b < 2 - 2a - 4\sqrt { - a}$, with $-1 < a < 0$. The
condition $a < 0$ provides  stability against continuum growth,
and the inequality
 $a > -1$ means that $\sigma$ cannot exceed the SAM
 depth. The positive [``$+$''sign in the first equation in Eq. (\ref{sol2})] and
 negative (``$-$''sign) branches of
 the solution ($\ref{sol2}$) coincide along the curve $
P\left( 0 \right) = {{3\left[ {2\left( {1 + a} \right) - b} \right]}
\mathord{\left/
 {\vphantom {{3\left[ {2\left( {1 + a} \right) - b} \right]} {4\varsigma b}}} \right.
 \kern-\nulldelimiterspace} {4 b \varsigma}} = {{3\left( {\sqrt {2b}  - b} \right)} \mathord{\left/
 {\vphantom {{3\left( {\sqrt {2b}  - b} \right)} {2b\varsigma }}} \right.
 \kern-\nulldelimiterspace} {2b\varsigma }}$.

Eq. (\ref{sol2}) demonstrates an existence of two control parameters
defining the CSP: $a$ and $b$. Physical meaning of the former one is
contribution of the saturated net-gain $\sigma$ (i.e. saturated gain
minus unsaturable loss) regarding the saturable loss ($\mu$ is the
maximum saturable loss coefficient, i.e. the SAM depth). The
$\mu$-parameter is a few or fraction of percents for the solid-state
CPO, but can be substantially larger for the ANDi oscillator (see
Section \ref{s3}). The $a$-parameter is not free, in fact, because
it depends on the pulse energy and, thereby, on the $b$-parameter
(see below).

Physical meaning of the $b$-parameter is closely related to the
mechanism of CSP formation \cite{proctor,haus2,oe}: i) phase
variation of the propagating chirped pulse can be balanced in the
presence of positive GDD, and ii) spreading of the chirped pulse can
be balanced by spectral filtering. In the cubic nonlinear version of
Eq. (\ref{GL}), the CSP chirp ($\psi \gg$1) is

\begin{equation} \label{chirp_sol}
\psi ^2  \approx {{\gamma P_0 T^2} \mathord{\left/
 {\vphantom {{\gamma P_0 } {\beta }}} \right.
 \kern-\nulldelimiterspace} {\beta }}.
\end{equation}

\noindent On the one hand, the peak power $P_0$ relates to the SAM
saturation power, that is $P_0 \propto \varsigma^{-1}$. On the other
hand, $\psi^2 / T^2 \propto \Delta^2$. Since the CSP spreading is
balanced by spectral filtering, one can roughly assume $\Delta^2
\propto 1/\alpha$. As a result of these rough estimations and Eq.
(\ref{chirp_sol}), one can see that the ratio $b \equiv \gamma
\alpha/\beta \varsigma \mu = const \cong 1$ represents a balance of
factors providing the CSP existence.

Returning to Eq. (\ref{sol2}), its integration results in the
implicit expression for the CSP profile

\begin{widetext}
\begin{equation}
t = \frac{3} {{\alpha \varsigma \Delta \Xi \left( {\Xi ^2  - \Delta
^2 } \right)}}\left\{ {\left[ {\left( {\Xi ^2  - \Delta ^2 }
\right)\beta \varsigma  - \gamma } \right]\Delta
\operatorname{arctanh} \left( {\frac{\Omega } {\Xi }} \right) +
\gamma \Xi \operatorname{arctanh} \left( {\frac{\Omega } {\Delta }}
\right)} \right\}\;, \label{sol3}
\end{equation}
\end{widetext}

\noindent where $ \Omega  =  \pm \sqrt {\Delta ^2  - {{\gamma P(t)}
\mathord{\left/
 {\vphantom {{\gamma P(t)} \beta }} \right.
 \kern-\nulldelimiterspace} \beta }}
$ from Eq. (\ref{sol1}). Equation (\ref{sol3}) allows the spectral
chirp to be expressed as

\begin{equation}\label{chirp}
\Psi  \equiv \frac{{d^2 \phi \left( \omega  \right)}} {{d\omega ^2
}} = \frac{{3\beta }} {{2\alpha }}\frac{{\Delta ^2  - \omega ^2  +
{\gamma  \mathord{\left/
 {\vphantom {\gamma  {\beta \varsigma }}} \right.
 \kern-\nulldelimiterspace} {\beta \varsigma }}}}
{{\left( {\Xi ^2  - \omega ^2 } \right)\left( {\Delta ^2  - \omega
^2 } \right)}}.
\end{equation}

\noindent The frequency dependence of $\Psi$ defines the CSP
compressibility \cite{kalash2,kalash3}: parts of the spectrum where
the chirp is strongly frequency-dependent, belong to the pulse
satellites after pulse compression. The flatness of the chirp in the
vicinity of $\omega=$0 enhances as the stability border $\sigma=$0
is approached. In contrast to the case of Ref. ~\cite{kalash1}, the
chirp is always minimum at the central frequency $\omega= $0.

The next step is to assume that $\phi(t)$ is a rapidly varying
function for the CSP \cite{kalash1,kalash2}. The stationary-phase method of
\cite{kalash2,sp} allows one to  express the
spectral power from the first of Eqs. (\ref{sol1}):

\begin{eqnarray} \label{sol4}
\begin{gathered}
  p\left( \omega  \right) \equiv \left| {\int\limits_{ - \infty }^\infty
  {dt\sqrt {P\left( t \right)} e^{i\phi \left( t \right) - i\omega t} } } \right|^2  \approx  \hfill \\
  \frac{{6\pi \beta ^2 }}
{{\gamma \alpha }}\frac{{\Delta ^2  - \omega ^2  + {\gamma
\mathord{\left/
 {\vphantom {\gamma  {\varsigma \beta }}} \right.
 \kern-\nulldelimiterspace} {\varsigma \beta }}}}
{{\Xi ^2  - \omega ^2 }}\Theta \left( {\Pi ^2  - \omega ^2 } \right) \hfill, \\
\end{gathered}
\end{eqnarray}

\noindent where $\Theta (x)$ is the Heaviside function and $\Pi=\min
\left\{ {\Delta ,\Xi } \right\}$ is the least of $\Delta$ and $\Xi$.
Equation (\ref{sol4}) demonstrates that the CSP has the spectrum
truncated at $\pm \Pi$ (i.e. $|\omega|<\Pi$).

Integration of Eq. (\ref{sol4}) allows the pulse
energy to be  expressed as

\begin{eqnarray} \label{sol5}
\begin{gathered}
  E \equiv \int\limits_{ - \infty }^\infty  {P\left( t \right)dt}
  \approx \int\limits_{ - \Delta }^\Delta  {p\left( \omega  \right)} \frac{{d\omega }}
{{2\pi }} \times  \hfill \\
  \frac{{6\beta ^2 \Delta }}
{{\alpha \gamma }}\left[ {1 - \frac{{\left( {\Xi ^2  - \Delta ^2  -
{\gamma  \mathord{\left/
 {\vphantom {\gamma  {\beta \varsigma }}} \right.
 \kern-\nulldelimiterspace} {\beta \varsigma }}} \right)\operatorname{arctanh} \left( {\frac{\Delta }
{\Xi }} \right)}}
{{\Delta \Xi }}} \right]. \hfill \\
\end{gathered}
\end{eqnarray}

\noindent It is clear from Eq. (\ref{sol5}) that the truncation
parameter $\Pi$ is equal to $\Delta$, i.e. $\Delta < \Xi$. In contrast
to the cubic-quintic CGLE, whose solution is the truncated Lorentz
function in the spectral domain \cite{kalash1}, the spectrum in our case
is parabolic-top ($\Delta < \Xi$): i) convex if $\alpha \Delta^2 < 3
(\mu+\sigma)$, ii) flat-top if $\alpha \Delta^2 = 3 (\mu+\sigma)$,
and iii) concave if $\alpha \Delta^2 > 3 (\mu+\sigma)$.

In an oscillator, the $\sigma$-($a$-)parameter is energy-dependent
owing to saturation of the gain $g$. The simplest law of saturation
is $g = {{g_0 } \mathord{\left/
 {\vphantom {{g_0 } {\left( {1 + {E \mathord{\left/
 {\vphantom {E {E_s }}} \right.
 \kern-\nulldelimiterspace} {E_s }}} \right)}}} \right.
 \kern-\nulldelimiterspace} {\left( {1 + {E \mathord{\left/
 {\vphantom {E {E_s }}} \right.
 \kern-\nulldelimiterspace} {E_s }}} \right)}}$ \cite{haus3},
  which is valid for the case if the active medium is small in comparison with the confocal length of
 laser beam ($g_0$ is the gain for a small signal and
 $E_s$ is the saturation energy). Such a law is typical for the fiber and solid-state thin-disk
 oscillators. Reverse relation between the confocal and active
 medium lengthes results in $
g = {{g_0 } \mathord{\left/
 {\vphantom {{g_0 } {\sqrt {1 + {E \mathord{\left/
 {\vphantom {E {E_s }}} \right.
 \kern-\nulldelimiterspace} {E_s }}} }}} \right.
 \kern-\nulldelimiterspace} {\sqrt {1 + {E \mathord{\left/
 {\vphantom {E {E_s }}} \right.
 \kern-\nulldelimiterspace} {E_s }}} }}
$ \cite{kalash3}. These expressions can be expanded in
 the vicinity of $\sigma=0$

 \begin{equation}\label{sol6}
\sigma \left( E \right) \approx \delta \left( {E - E^* } \right).
 \end{equation}

\noindent The $E^*$-parameter is the energy defined as the averaged
power of a free-running oscillator multiplied by the cavity period
$T_{cav}$. $ E^* = {{\left( {g\left( 0 \right) - l - \mu }
\right)E_s } \mathord{\left/
 {\vphantom {{\left( {g\left( 0 \right)  - l - \mu } \right)E_s } {\left( {l + \mu } \right)}}} \right.
 \kern-\nulldelimiterspace} {\left( {l + \mu } \right)}}
$ for the first law of gain saturation (see above and Ref.
\cite{haus3}). $ \delta \equiv \left. {{{d\sigma } \mathord{\left/
 {\vphantom {{d\sigma } {dE}}} \right.
 \kern-\nulldelimiterspace} {dE}}} \right|_{E = E^* }
$ [i.e. $ \delta  =  - {{\left( {l + \mu } \right)^2 }
\mathord{\left/
 {\vphantom {{\left( {l + \mu } \right)^2 } {g\left( 0 \right)E_s }}} \right.
 \kern-\nulldelimiterspace} {g\left( 0 \right)E_s }}
$] is the parameter permitting the gain saturation.  For the
considered law of gain saturation, one has $\rho\equiv E^* \delta =
{{-g\left( 0 \right)\left( {\xi - 1} \right)} \mathord{\left/
 {\vphantom {{g\left( 0 \right)\left( {\xi  - 1} \right)} {\xi ^2 }}} \right.
 \kern-\nulldelimiterspace} {\xi ^2 }} $, and $\xi \equiv g(0)/(l + \mu) \approx
 g(0)/l$ is the pump-to-threshold ratio.

The normalizations presented in Table \ref{table} reduce the
three-parametrical space of Eq. (\ref{GL}) to a two-parametrical one
$\left( {a ,b} \right)$ for the CSP. The resulting dimensionless
equations are shown in Table \ref{table2}. Thus, the CSP is easily
traceable on a two-dimensional plane (``master diagram'') as in  the
case of the nonlinear cubic-quintic CGLE
\cite{kalash1,kalash2,kalashn}.

\begin{table}
\caption{\label{table}Relations between dimensional and
dimensionless quantities}
\begin{ruledtabular}
\begin{tabular}{lp{2in}}
$E$ & $E' \beta^2 \sqrt{\mu}/\gamma \alpha \sqrt{\alpha}$\\
$E^*$ & $E^{*'} \beta^2 \sqrt{\mu}/\gamma \alpha \sqrt{\alpha}$\\
$P$ & $P'/ \varsigma$\\
$p$ & $p' \beta^2/\alpha \gamma$\\
$\omega$ & $\omega' \sqrt{\mu/\alpha}$\\
$\Omega$ & $\Omega' \sqrt{\mu/\alpha}$\\
$\Delta$ & $\Delta' \sqrt{\mu/\alpha}$\\
$\Xi$ & $\Xi' \sqrt{\mu/\alpha}$\\
$\Psi$ & $\Psi' \beta/\mu$
\end{tabular}
\end{ruledtabular}
\end{table}

\begin{center}
\begin{table}
\caption{\label{table2}Dimensionless equations for the CSP
parameters ($a\equiv \sigma/\mu$, $b \equiv \alpha \gamma/\beta
\varsigma \mu$)}
\begin{ruledtabular}
\begin{tabular}{c}

$P'\left( 0 \right) = \frac{3} {{4b}}\left[ {2\left( {1 +
a} \right)
- b \pm \sqrt \Upsilon  } \right]$\\
$\Delta'^2 = \frac{3} {4}\left[ {2\left( {1 + a} \right) - b \pm
\sqrt
\Upsilon  } \right]$\\
$\Upsilon  = \left( {b - 2} \right)^2  + 4a\left( {2 + a + b}
\right)$\\
$
\Xi '^2  = \frac{2} {3}\Delta '^2  + b + a + 1$\\
 $\Psi ' = \frac{3}
{2}{{\left( {\Delta '^2  - \omega '^2  + b} \right)} \mathord{\left/
 {\vphantom {{\left( {\Delta '^2  - \omega '^2  + b} \right)} {\left( {\Xi '^2  - \omega '^2 } \right)\left( {\Delta '^2  - \omega '^2 } \right)}}} \right.
 \kern-\nulldelimiterspace} {\left( {\Xi '^2  - \omega '^2 } \right)\left( {\Delta '^2  - \omega '^2 } \right)}}$\\
$p' = 6\pi {{\left( {\Delta '^2  - \omega '^2  + b} \right)\Theta
\left( {\Delta '^2  - \omega '^2 } \right)} \mathord{\left/
 {\vphantom {{\left( {\Delta '^2  - \omega '^2  + b} \right)\Theta \left( {\Delta '^2  - \omega '^2 } \right)} {\left( {\Xi '^2  - \omega '^2 } \right)}}} \right.
 \kern-\nulldelimiterspace} {\left( {\Xi '^2  - \omega '^2 }
 \right)}}
$\\
$E' = 6 \Delta' \left[ {1 - {{\left( {\Xi '^2  - \Delta '^2  - b}
\right)\operatorname{arctanh} \left( {\frac{{\Delta '}} {{\Xi '}}}
\right)} \mathord{\left/
 {\vphantom {{\left( {\Xi '^2  - \Delta '^2  - b} \right)\operatorname{arctanh} \left( {\frac{{\Delta '}}
{{\Xi '}}} \right)} {\Delta '\Xi '}}} \right.
 \kern-\nulldelimiterspace} {\Delta '\Xi '}}} \right]
$
\end{tabular}
\end{ruledtabular}
\end{table}
\end{center}

\section{\label{s2}Chirped-pulse oscillators: comparative analysis}

\begin{figure}
\includegraphics[width=9.5cm]{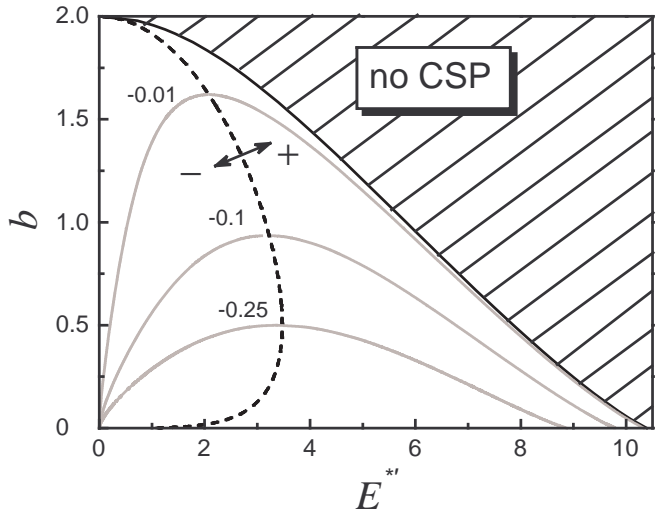}
\caption{\label{fig1} The master diagram. The solid black curve is
the CSP stability border. The dashed curve is the border between the
$+$ and $-$ branches of CSP for $\rho/\mu=$17. The gray curves are
the isogains for $a$ = -0.01, -0.1 and -0.25 (superscribed).}
\end{figure}

The master diagram on the plane $\left( b ,\,E^* \right)$ is shown
in Fig. \ref{fig1}. The black curve is the stability threshold
$\sigma= $0 (that is,  $E=E^*$). Above this curve, the CSP does not
exist (hatched region). Along this curve, the dimensionless pulse
parameters (see Tables \ref{table},\ref{table2}) are

\begin{equation}\label{sol7}
\begin{gathered}
  0<b<2,\,\, P'(0)  = \Delta'^2/b  = 3\left( {\frac{1}{b} - \frac{1 }
{2}} \right),\,\,\Xi'^2  = 3, \hfill \\
  E^{*'} = 6\sqrt {3\left( {1 - \frac{b }
{2}} \right)} \left[ {1 - \frac{{b \operatorname{arctanh} \left(
{\sqrt {1 - \frac{b } {2}} } \right)}} {{6\sqrt {1 - \frac{b }
{2}} }}} \right]. \hfill \\
\end{gathered}
\end{equation}

\noindent The dashed curve in Fig. \ref{fig1} shows the border
between the positive ($+$) and negative ($-$) branches of Eq.
(\ref{sol2}).

The CSPs providing a constant value of the saturated net-gain
parameter $\sigma$ correspond to the gray curves in Fig. \ref{fig1}
(so-called, isogain curves). The isogain $\sigma=0$ is the stability
threshold (solid black curve), and only three nonzero isogain curves
are shown (the corresponding values of $a$ are superscribed in Fig.
\ref{fig1}).

\begin{figure}
\includegraphics[width=9.5cm]{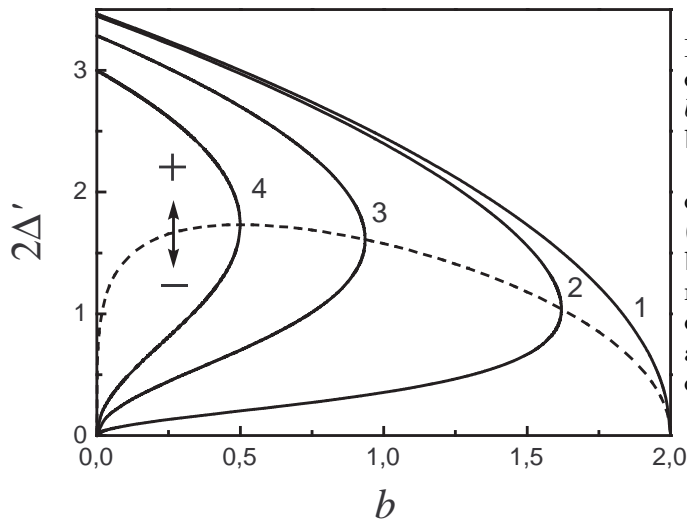}
\caption{\label{fig2} The spectral widths in the isogain curves of
Eq. (\ref{sol2}) ($a=$ 0 (1), -0.01 (2), -0.1 (3) and -0.25 (4)).
The upper (lower) branches of the curves correspond to the
$+$($-$)branches of Eq. (\ref{sol2}); the dashed curve is the border
between branches of CSP.}
\end{figure}

Since $\sigma$ is a function of $E/E^*$, the isogain curves explain
the meaning of the $+$- and $-$ branches (see also
\cite{kalash1,kalashn}). The $+$ branch corresponds to the
energy-scalable CSP. This means that $E$ grows $\propto E^*$ along
the isogain faster than the $b$-parameter ($b \propto 1/\beta$)
decreases with the dispersion $\beta$. That is,  keeping in such an
isogain with the energy scaling (note that $E^{*'}\propto
E^*/\beta^2$) needs a comparatively slow GDD increase. As a result,
the SPM increases faster than the spectrum degrades with the GDD.
That is the spectrum broadens (see Fig. \ref{fig2}, where
$b$-decrease along a $+$ isogain corresponds to  the $E^{*'}$-growth
in  Fig. \ref{fig1}). The spectra are broadest on the stability
border.

For the $+$ branch, the spectrum narrows with $\alpha$ (Fig.
\ref{fig2}; $b \propto \alpha$). This can be explained as a result
of the $E^*$-decrease, which is necessary for keeping in the isogain
($ E^{*'}   \propto \alpha ^{3/2} E^*$ ) (Fig. \ref{fig1}). That is,
the SPM contribution decreases and the spectrum narrows.

The $-$ branch corresponds to the energy-unscalable CSP. This means
that $b$ scaling ($b \propto1/\beta$) weakly affects $E^*$
($E^{*'}\propto E^*/\beta^2$) (Fig. \ref{fig1}). Thus, the energy
remains almost constant along this isogain when the GDD scales.
Certainly, energy scaling is possible as well. However, that is not
isogain process for this branch of the CSP. For the $-$ branch the
spectrum narrows with the $b$-decrease ($b \propto 1/\beta$) due to
growth of the GDD contribution, which stretches the pulse when the
energy remains almost constant (Fig. \ref{fig2}).

When $E^*$ changes weakly along the isogain corresponding to the $-$
branch, the spectrum broadens with $\alpha$ ($b \propto
\alpha$)(Fig. \ref{fig2}). The explanation is that the growth of
spectral filtering enhances the cutoff of red (blue)-shifted
spectral components located on the pulse front (tail). The spectral
shift at the pulse edges is a consequence of chirp
\cite{proctor,haus2}. The growth of cutoff shortens the CSP and, for
a fixed energy, $P_0$ increases. Since $P_0 \propto \Delta^2$, the
spectrum broadens.

One can additionally clarify the division into the $+$ and $-$
branches by considering the nonlinear cubic limit of Eq. (\ref{GL}).
Such a limit describes a low-energy CPO \cite{haus2}. In this case,
the exact CSP solution is $\propto$ sech($t/T$)$^{1+i \psi}$, and
its spectral profile can be expressed through beta functions. By the
method described in the previous section it is readily found that

\begin{equation}\label{cubic}
\begin{gathered}
  \gamma P\left( t \right) = \beta \Delta^2 \left( {1  - \tanh^2 \left[ {{{\Delta t\xi \left( {1 + b} \right)} \mathord{\left/
 {\vphantom {{\Delta t\xi \left( {1 + b} \right)} {3\gamma }}} \right.
 \kern-\nulldelimiterspace} {3\gamma }}} \right]} \right), \hfill \\
  \alpha \Delta ^2  = \frac{{3\sigma b}}
{{b - 2}}, \hfill \\
  p\left( \omega  \right) = \frac{{6\pi \beta }}
{{\left( {1 + b} \right)\xi }} \Theta \left( {\Delta ^2  - \omega ^2
} \right)
. \hfill \\
\end{gathered}
\end{equation}

\noindent Here $\xi \equiv \mu \varsigma$. Equation (\ref{cubic})
demonstrates that i) CSP exists only for $\sigma<$0, ii) its
spectral width increases with $b$ and $|\sigma|$. Comparison with
Fig. \ref{fig2} suggests that such a behavior corresponds to the $-$
branch.

On the basis of the model developed, three types of CPOs will be
considered: i) all-normal-dispersion (ANDi) fiber oscillator (like
that in Ref.~\cite{wise2}), ii) broadband (e.g. Ti:sapphire
\cite{naumov,kalash1} or Cr:YAG \cite{sorokin}) and iii) narrow-band
(e.g. Yb-doped \cite{morgner2,morgner3,morgner4}) solid-state
oscillators. Representation of these on the master diagram affords a
means of better understanding and controlling of CPO (see also Ref.
\cite{kalashn}).

\subsection{\label{s21}ANDi fiber oscillator}

In Section \ref{s1}, the CSP solution of the generalized nonlinear
CGLE was obtained in the
limits $\alpha\ll\beta$ 
and $\beta\ll T^2$. The master parameter $b \equiv \alpha
\gamma/\beta \varsigma \mu$ controlling the CSP is defined by
the relative but not absolute contributions from the dissipative and
non-dissipative factors of the CGLE. This allows a unified standpoint on
CPOs with parameters which vary within a broad range.

\begin{figure}
\includegraphics[width=9.5cm]{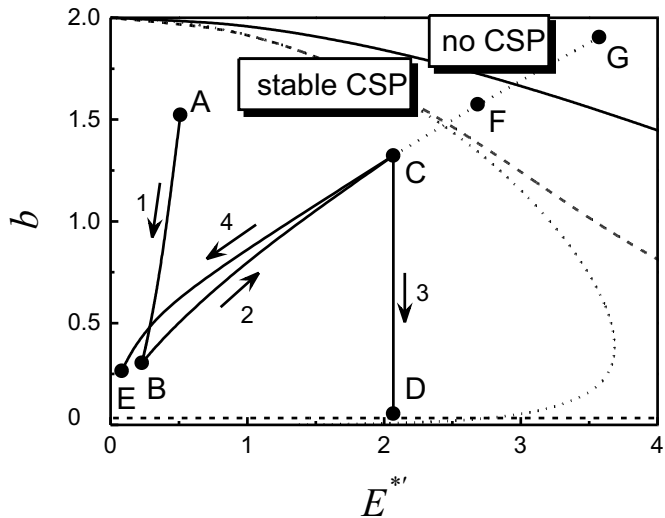}
\caption{\label{fig3} Sector of master diagram (Fig. \ref{fig1})
representing the ANDi fiber oscillators. Parameters corresponding to
the \emph{A} -- \emph{G} points are given in Table \ref{table3}. The
solid curve is the stability border; the dotted and dashed curves
are the borders between the $+$ and $-$ branches of solution
(\ref{sol2}) for $\rho/\mu= $3.8 and 0.76, respectively (i.e.,
$\mu=$0.1 and 0.5 for $\rho=$0.38, respectively).}
\end{figure}

Figure \ref{fig3} shows the sector of the master diagram covering
the ANDi fiber oscillator parameters (see Table \ref{table3}; the
estimations for the parameters are based on Refs.
\cite{wise2,chong}). Let us start at point $A$ (Fig. \ref{fig3} and
Table \ref{table3}) corresponding to a typical set of
ANDi-oscillator parameters but with the comparatively small SAM
depth $\mu$. Although the GDD value is large in comparison with that
in a CPO, the spectral filter bandwith (25 nm) is small. As a
result, the excess of the ratio $\beta/\alpha$ over that for a
Ti:sapphire CPO is only tenfold (see below and Ref. \cite{kalash1}).
Simultaneously, an excess of the ratio $\gamma/\mu \varsigma$ over
that for a Ti:sapphire CPO is tenfold as well. As a result, the
$b$-parameter is $\simeq$1 and, dynamically, there is no substantial
distinction in kind between the ANDi fiber and the solid-state CPOs.
One difference is that a large GDD and a comparatively small $E^*$
shift the operational point into the $-$ branch region (Fig.
\ref{fig3}).

\begin{table}[h]
\begin{center}
\caption{\label{table3}Parameters of ANDi oscillator corresponding
to Fig. \ref{fig3}. $E^*=$12 nJ, $\gamma=$0.014 W$^{-1}$. The spectrum
is centered at $\approx$1 $\mu$m.}
\begin{tabular}{|l|l|l|l|l|l|l|l|}
\hline
& $A$ & $B$ & $C$  & $D$ & $E$ & $F$ & $G$\\
\hline $\alpha$ (fs$^2$)& $450$ & $450$ & $2000$  & $2000$ & $2000$ & $2330$ & $2820$\\
\hline
$\beta$ (ps$^2$)& $0.1$ & $0.1$ & $0.1$  & $0.1$ & $0.5$ & $0.1$ & $0.1$\\
\hline
$\mu$ & $0.1$ & $0.5$ & $0.5$  & $0.5$ & $0.5$ & $0.5$ & $0.5$\\
\hline
$\varsigma$ (kW$^{-1}$) & $0.42$ & $0.42$ & $0.42$  & $10$ & $0.42$ & $0.42$ & $0.42$\\
\hline
\end{tabular}
\end{center}
\end{table}

\begin{figure}
\includegraphics[width=9.5cm]{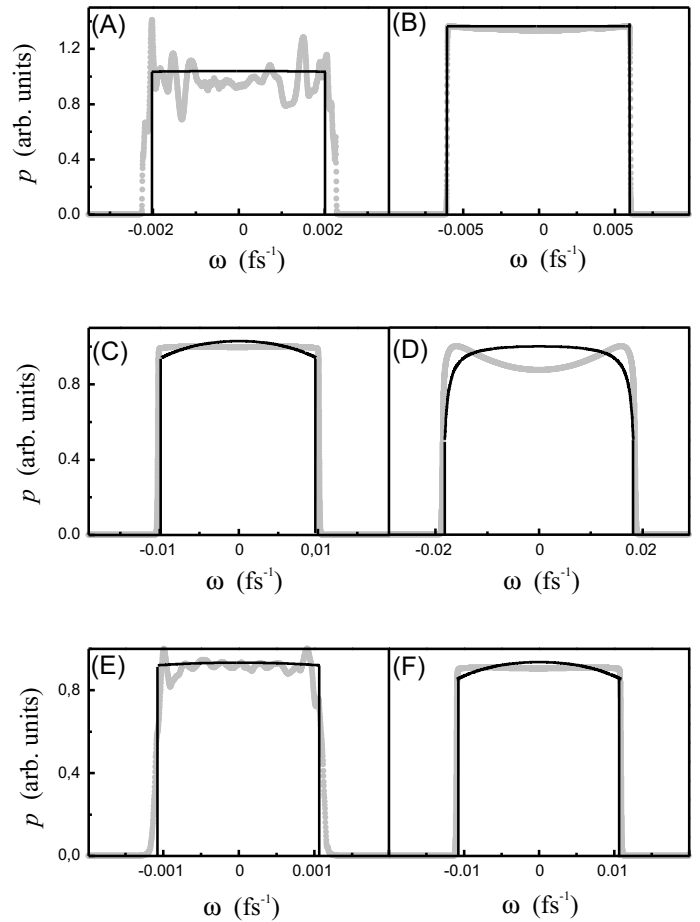}
\caption{\label{fig4} Numerical (gray circles) and analytical (solid
black curves) spectra of ANDi fiber oscillators with the parameters
defined in Table \ref{table3}; $\rho=$0.38.}
\end{figure}

\emph{A} of Fig. \ref{fig4} shows the spectrum of the numerical
solution of Eq. (\ref{GL}) (gray circles) and the analytical profile
(\ref{sol4}) (solid curve) corresponding to  point $A$ in Fig.
\ref{fig3}. The analytical profile reproduces the averaged numerical
one. As can be seen, the numerical solution is strongly perturbed.
This effect was first reported in Ref. \cite{kalash4} and attributed
to excitation of the solitonic internal modes. Such modes grow with
the GDD and are excited when the value $\beta/\alpha$ is large.

A way of suppressing such a perturbation is to increase the SAM depth
$\mu$ (transition 1 from $A$ to $B$ in Fig. \ref{fig3}). The
$\mu$-growth is not the isogain process, as the $|a|$-parameter
increases. As a result, the spectrum broadens (Fig. \ref{fig4},
\emph{B}) according to  the analytical model (transition from the lower
branch of curve 2 to that of 4 in Fig. \ref{fig2}).

An important factor governing the CPO is spectral filtering, since
the pulse lengthening due to GDD has to be compensated by its
shortening  owing to frequency filtering of the chirped pulse
\cite{haus2}. In the framework of the model under consideration, the
filter band narrowing can be illustrated by transition 2 to
point $C$ in Fig. \ref{fig3}. Such a transition results in almost
isogain growth of the $b$-parameter. Hence, for the ``$-$''branch
of the CSP (Fig. \ref{fig2}) the spectrum broadens (Fig. \ref{fig4},
\emph{C}) \cite{wise2}. The further filter band narrowing transforms
the pulse into the ``$+$'' branch of the CSP (Fig. \ref{fig4}, $F$).
Further growth of $b \propto \alpha$ due to  filter band narrowing
destabilizes the pulse ($\sigma> $0 at point $G$).

The next important factor is the inverse power of the loss
saturation (the SAM parameter $\varsigma$). Its growth (transition 3
from $C$ to $D$ in Fig. \ref{fig3}) corresponds to the $b$-decrease.
First, this is not an isogain process, i.e. the $|a|$-parameter
increases, which broadens the spectrum (Fig. \ref{fig2}). Second,
the spectrum narrows with the $b$-decrease for the $-$ branch of the
CSP (Fig. \ref{fig2}). In our example, point $D$ is located in the
vicinity of the border between the $-$ and $+$ branches of the CSP.
In accordance with Figure \ref{fig2}, this means that the spectrum
broadens owing to the $|a|$-parameter growth (Fig. \ref{fig4},
\emph{D}). \emph{D} of Fig. \ref{fig4} demonstrates that, unlike the
analytical profile, the numerical one is distinctly concave. This
issue will be discussed in Sec. \ref{s3}.

The characteristic of the ANDi fiber oscillator is that it is
possible to vary the positive GDD within a wide range \cite{wise2}.
The GDD growth decreases the $b$-parameter and, as a result, narrows
the spectrum of the CSP relating to the $-$ branch (Figs. \ref{fig2}
and \ref{fig4}, \emph{E}) \cite{wise2}. Such a conclusion is valid
for both isogain and non-isogain variation. In the latter case, the
$|a|$-parameter decreases with the GDD growth for a fixed $E^*$,
which enhances the narrowing of the spectrum for the ``$-$'' branch
(Fig. \ref{fig2}). Solitonic internal modes again occur
\cite{kalash4} and the spectrum becomes perturbed (Fig. \ref{fig4},
\emph{E}).

\subsection{\label{s22}Broadband solid-state CPO}

As was pointed out, there are two distinctive differences between
the ANDi fiber oscillator and the solid-state CPO: the former has
substantially larger GDD and SPM. Nevertheless, both oscillators can
be described from a unified standpoint because their properties are
defined by only two dimensionless parameters,  $b$ and $E^*$. From
this point of view, the main difference between them is that the
ANDi oscillator belongs mainly to the $-$ branch of the CSP, whereas
the CPO belongs to the $+$ branch. It should be noted that this
statement need not be considered categorically, because the growth
of SAM and spectral filtering shifts the operational point of an
ANDi oscillator into the $+$ branch region (see points $D$ and $F$
in Fig. \ref{fig3}).

Let us consider a broadband (Ti:sapphire) CPO with the parameters
presented in Table \ref{table4}. The mode-locking is provided by
SESAM with the inverse saturation power $\varsigma$, which
corresponds to a  saturation energy fluence of 100 $\mu$J/cm$^2$,
a relaxation time of 0.5 ps and a mode radius of 100 $\mu$m.

\begin{table}[h]
\begin{center}
\caption{\label{table4}Parameters of a Ti:sapphire CPO corresponding
to Fig. \ref{fig5}. $\alpha=$2.5 fs$^2$, $\gamma=$4.55 MW$^{-1}$,
$\varsigma$=16 MW$^{-1}$. The spectrum is centered at $\approx$ 0.8
$\mu$m.}
\begin{tabular}{|l|l|l|l|l|l|l|l|}
\hline
& $A$ & $B$ & $C$  & $D$\\
\hline
$\beta$ (fs$^2$)& $160$ & $160$ & $160$  & $180$ \\
\hline
$\mu$ & $0.01$ & $0.02$ & $0.02$  & $0.02$\\
\hline
$E^*$ (nJ) & 1120 & 1120 & 2240  & 2240 \\
\hline
\end{tabular}
\end{center}
\end{table}

\begin{figure}
\includegraphics[width=9.5cm]{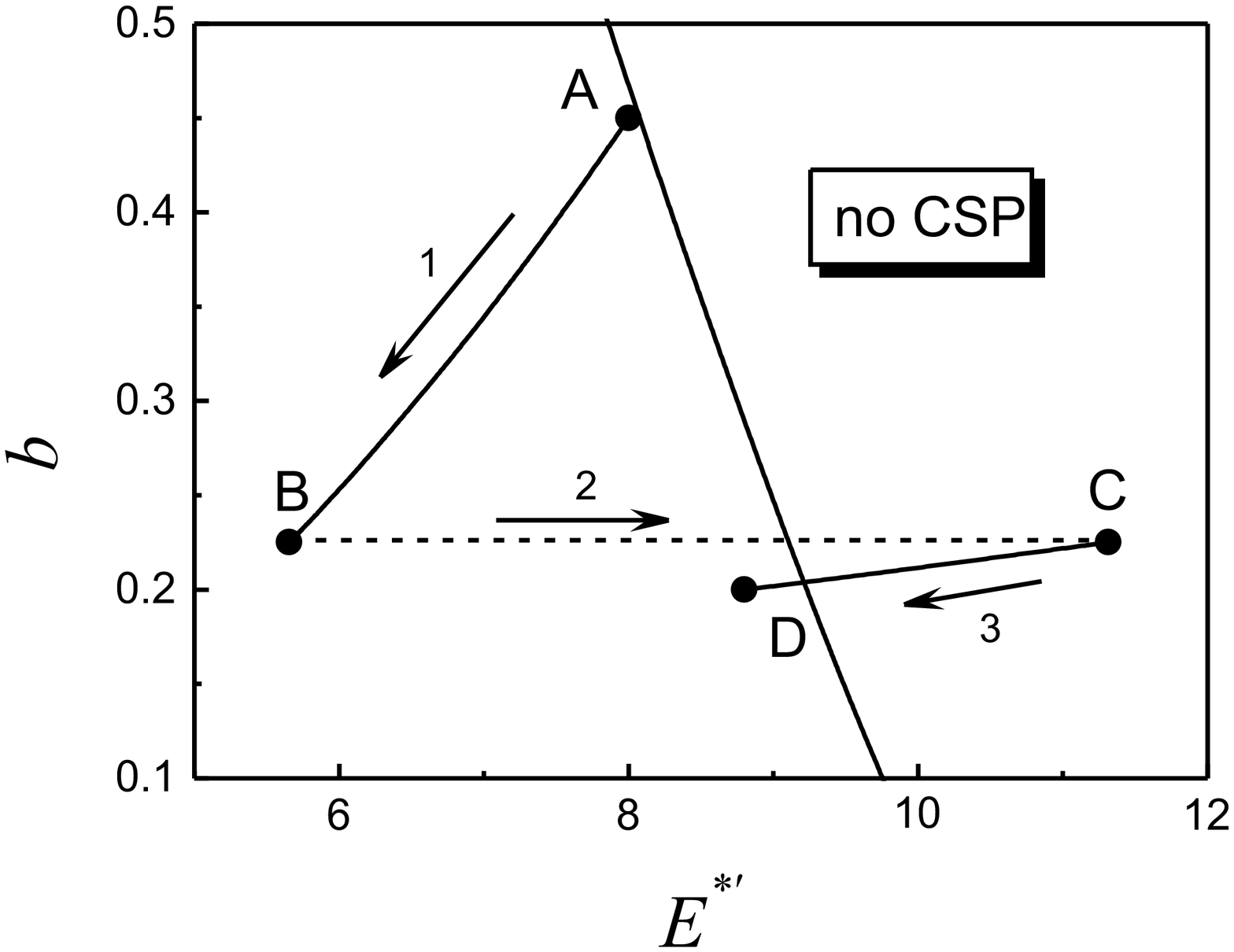}
\caption{\label{fig5} Sector of master diagram (Fig. \ref{fig1})
representing the broadband solid-state CPOs. Parameters
corresponding to points  \emph{A} -- \emph{D} are given in Table
\ref{table4}. The solid curve is the stability border.}
\end{figure}

\begin{figure}
\includegraphics[width=10cm]{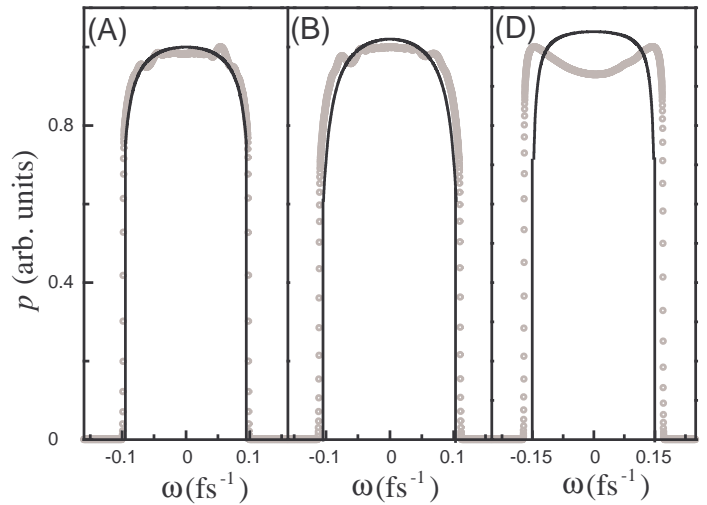}
\caption{\label{fig6} Numerical (gray circles) and analytical (solid
black curves) spectra of CPO oscillators with the parameters defined
in Table \ref{table4}; $\rho=$0.17.}
\end{figure}

The operational point $A$ on the stability border (Fig. \ref{fig5})
corresponds to the minimum GDD and the broadest (for a given
set of parameters) spectrum ($\approx$70 nm, see Fig. \ref{fig6}
$A$).

An attempt to increase the energy would destabilize the CSP.
Therefore, at first it is useful to increase the modulation depth
(transition 1 to from $A$ to $B$ in Fig. \ref{fig5}). This is not an
isogain process, and therefore the spectrum (Fig. \ref{fig6}, $B$)
does not change substantially in spite of the $b$-decrease (the
$b$-decrease broadens the $+$ branch spectrum, but the $\mid
a\mid$-growth narrows it, see Fig. \ref{fig2}).

The increase of the inverse saturation power $\varsigma$ (e.g. by
means of mode reduction) would be useful for subsequent
energy growth. However, as a rule, the SESAM operates in the
vicinity of the damage threshold, and so the $\varsigma$-growth
can be problematic.

The stability reserve obtained  (point $B$ in Fig. \ref{fig5})
allows energy scaling (transition 2 from $B$ to $C$ in Fig.
\ref{fig5}). Nevertheless, the twofold energy growth destabilizes
the CSP in our case, and  multipulsing occurs.

The way out is to increase the GDD (transition 3 from $C$ to $D$ in
Fig. \ref{fig5}), which shifts the operational point inside the
stability range. Since the $A \rightarrow D$ transition is an
isogain process following the $b$-decrease (curve with $a= $0 in
Fig. \ref{fig2}), the spectrum broadens (Fig. \ref{fig6},$D$).
Again, the spectrum becomes concave.

\begin{figure}
\includegraphics[width=9.5cm]{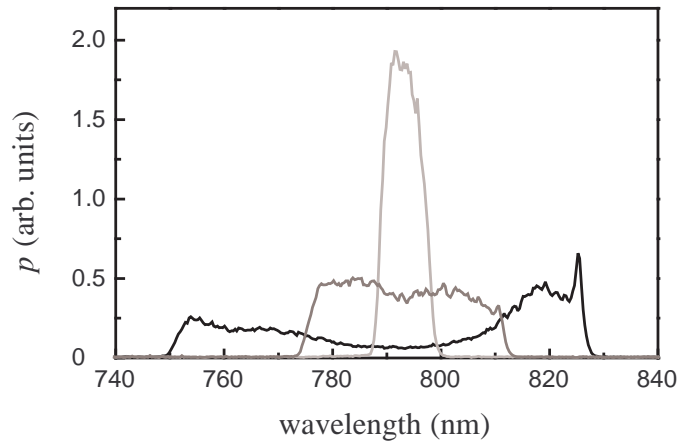}
\caption{\label{exp1} Experimental spectra from the Ti:sapphire CPO
corresponding to the growing positive net GDD (from the black via
gray to the light gray curves). Pump power 5.5 W; output power is of
1.25 W in the mode-locked regime.}
\end{figure}

It should be noted that, in an experiment, the main control
parameter used for oscillator stabilization at some fixed
$E^*$-level is the GDD value. The GDD growth (corresponding to the
$b$-decrease) is not an isogain process for the $+$ branch of the
CSP and is almost an isogain one for the $-$ branch. Figure
\ref{fig2} shows that the spectrum narrows in the latter case. For
the $+$ branch, the $|a|$-parameter increases with GDD, which
enhances the CSP stability against the continuum growth. Numerical
simulations demonstrate that the effect of the GDD growth on the CSP
spectrum in the case under consideration is its narrowing and
stretching of the CSP. Such a conclusion agrees with the experiment
\cite{naumov}. Figure \ref{exp1} shows the spectra from the
Ti:sapphire CPO corresponding to the positive net GDD growing step
by step. The spectra narrow with the GDD growth and reshape from
concave via flat-top to parabolic. Equation (\ref{sol4}) reproduces
the growth of convexity with narrowing of the spectrum. The concave
spectra appear in the numerical simulations, when the spectrum
becomes sufficiently broad (Figs. \ref{fig4}, \emph{D} and
\ref{fig6}, \emph{D}). In contrast to Ref. \cite{kalashn}, the
present analytical theory fails to reproduce this phenomenon (see
Section \ref{s3}).

\subsection{\label{s23}Narrowband solid-state CPO}

As  was found in Ref. \cite{kalash1}, there is a  limit to
CPO energy growth with resonator lengthening, because the modulation
depth $\mu$ has to increase $\propto T_{cav}/T_r$ (here $T_r$ is the
gain relaxation time). Since broadband solid-state active media
such as Ti:sapphire, Cr-doped zinc-chalcogenides, and Cr:YAG have
comparatively short gain relaxation times (a few microseconds), it
is preferable to use media with a long relaxation time,  such as
Yb-doped crystals. Moreover, the cavity length realized so far is
already in the MHz-range in terms of repetition rate, meaning that the
only scaleable parameter remains the power. Power scaling is
realizable in a Yb-doped thin-disk oscillators. For example, the
Yb:YAG oscillator operating in the NDR has exceeded the 13-$\mu$J
energy frontier \cite{neuhaus}. Such a regime requires a fair amount
of negative GDD ($\approx$-0.2 ps$^2$ in the case of Ref.
\cite{neuhaus}) and the pulse obtained is linearly incompressible.

It is interesting to consider the prospects of such a regime within
the PDR. The issue is that a Yb-doped medium has a comparatively
narrow gainband ($\alpha\approx$1000 fs$^2$) and it is not clear
\emph{a priori} that the CPO can operate at  GDD levels close to
$\alpha$. Nevertheless, Yb-doped CPOs have been demonstrated
experimentally \cite{morgner2,morgner3}. The CSPs obtained were
compressible down to $\approx$450 fs$^2$ and the positive net GDD
varied within the range  $\approx$250 -- 2250 fs$^2$.

Let us consider a Yb:YAG thin-disk CPO mode-locked by SESAM and
aiming at an intracavity pulse energy level $\approx$80 $\mu$J. At such
an energy level, an important factor is the SPM due to air filling
the resonator \cite{keller3}. Such a factor has to be taken into
account. Let the cavity length be 15 m, and the averaged mode
diameter  be equal to 2.4 mm. The Yb:YAG-disk thickness is 0.4 mm. The
SESAM saturation energy fluence is 100 $\mu$J/cm$^2$, its relaxation
time is 0.6 ps, and the mode diameter on SESAM is 1.2 mm. Other
parameters are presented in Table \ref{table5}.

\begin{table}[h]
\begin{center}
\caption{\label{table5}Parameters of a Yb:YAG CPO: $\alpha= $900
fs$^2$, $\varsigma= $0.53 MW$^{-1}$, $\mu= $0.005, $E^*= $80 $\mu$J.
The spectrum is centered at $\approx$1 $\mu$m.}
\begin{tabular}{|l|l|l|l|}
\hline
& $A$ & $B$ & $C$\\
\hline
$\beta$ (fs$^2$)& $2550$ & $700$ & $2900$\\
\hline
$\gamma$ (GW$^{-1}$) & $1.9$ & $0.15$ & $2.4$\\
\hline
\end{tabular}
\end{center}
\end{table}

\begin{figure}
\includegraphics[width=9.5cm]{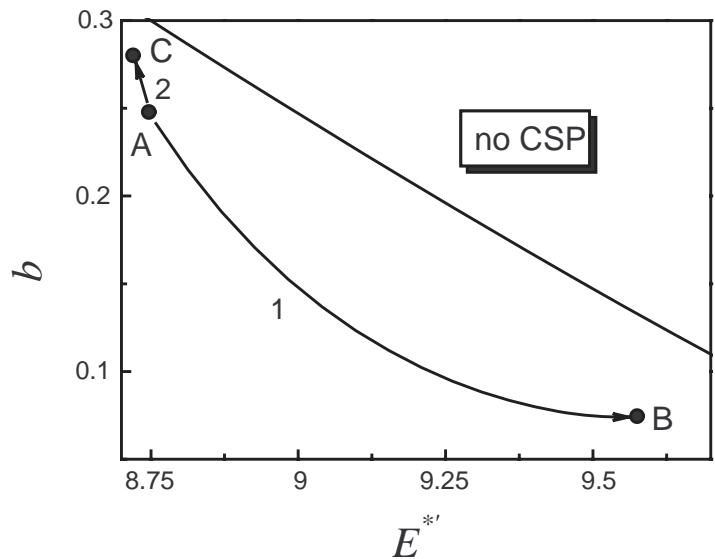}
\caption{\label{fig7} Sector of master diagram (Fig. \ref{fig1})
representing the narrowband solid-state CPOs. Parameters
corresponding to points \emph{A} -- \emph{C} are given in Table
\ref{table5}. The solid curve is the stability border.}
\end{figure}

\begin{figure}
\includegraphics[width=9.5cm]{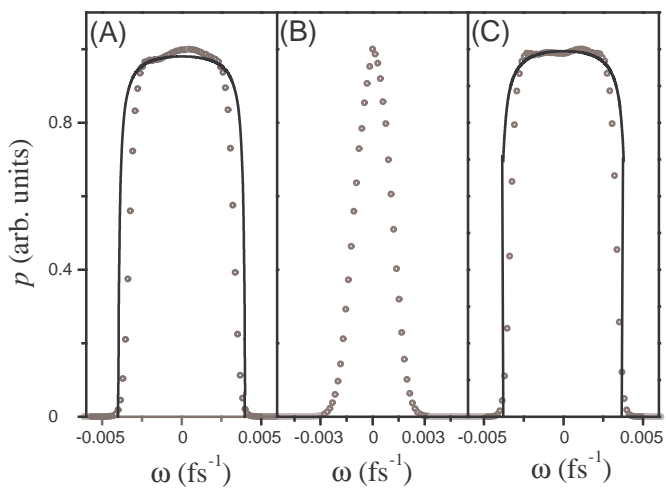}
\caption{\label{fig8} Numerical (circles) and analytical (curves)
spectra of CPO oscillators with the parameters defined in Table
\ref{table5}; $\rho=$0.05.}
\end{figure}

Point $A$ in Fig. \ref{fig7} corresponds to the narrowband CPO
operating in the vicinity of the stability border. The CPO resonator
is filled with air. The CSP belongs to the $+$ branch;  the
corresponding spectrum is shown in Fig. \ref{fig8}, \emph{A}. Its
width is $\approx$3.5 nm,  which allows compressing linearly down to
$\approx$300 fs. Although the ratio $\beta/\alpha$  is only 2.8, the
analytical profile provides quite precise fitting of the numerical
spectrum.

Violation of the assumption of $\beta \gg \alpha$ underlying the
analytical model results in  smoothed spectrum edges. Such
smoothing increases (Fig. \ref{fig8}, \emph{B})  when the resonator
becomes airless and only the active medium nonlinearity contributes
to the oscillator dynamics. Point $B$ in Fig. \ref{fig7}
corresponds to the neighborhood of the analytical stability border,
i.e. the almost minimum possible GDD value providing the  broadest
spectrum. Since $\beta < \alpha$, the analytical model is not valid,
although the pulse is chirped and remains almost fourfold
compressible (down to $\approx$800 fs).

\begin{figure}
\includegraphics[width=10cm]{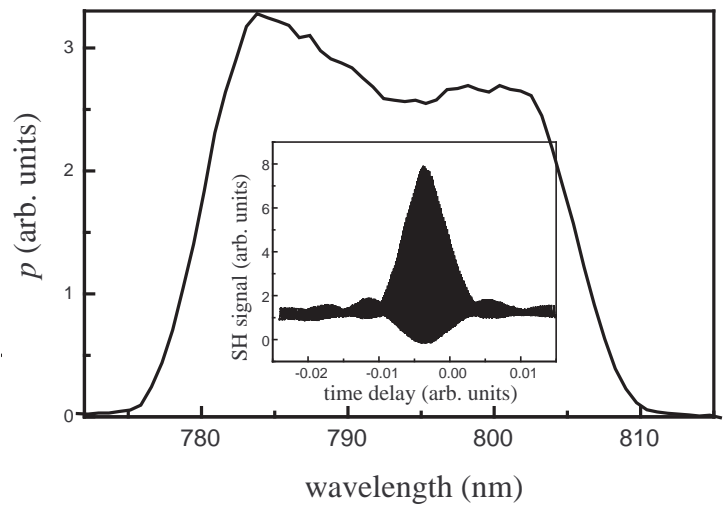}
\caption{\label{exp2} Experimental spectrum and autocorrelation
function (inset) from the Ti:sapphire CPO mode-locked by SESAM.
Repetition rate is of 60 MHz; intracavity energy is of 200 nJ.}
\end{figure}

The net-gainband narrowing (growth of $\alpha$ in comparison with
$\beta$) can also result from  spectral filtering produced by
comparatively narrowband SESAM. In this case, the $\alpha$-parameter
is defined by the SESAM bandwidth, and the approach of the
$\alpha$-parameter to the $\beta$-parameter results in smoothing of the spectrum
edges in a Ti:sapphire oscillator as well (Fig. \ref{exp2}).

Spectrum control of the narrowband CPO operating in the vicinity
of the stability border can be provided by inserting a plate (e.g., a
sapphire plate) introducing additional SPM (point $C$ in Fig.
\ref{fig7}; 0.1-cm sapphire plate). Figure \ref{fig8}, \emph{C}
demonstrates that the $b$-growth ($b \propto \gamma$) provides
excellent agreement between the numerical and analytical solutions.

It can be concluded that the absolute value of GDD required for
high-energy pulse stabilization is substantially lower in the PDR
than that in the NDR, so that one can avoid the resonator helium
filling or vacuumization of the resonator. Nevertheless,
vacuumization of the oscillator can provide the most direct way to
substantial energy growth. Let us consider an example with an
intracavity energy of  $\approx$0.8 mJ for a configuration
corresponding to point $C$ in Fig. \ref{fig7}. The scaling rules
$E^{*'}  = E^* \gamma \alpha ^{3/2} /\sqrt \mu \beta ^2$, $b =
\alpha \gamma /\beta \varsigma \mu$ (Table \ref{table}) demonstrate
that tenfold reduction of SPM (as a result of resonator
vacuumization) and SAM (as a result of, for instance, mode growth
and/or reduction of the saturation energy fluence) allow the system
to be kept at point $C$. This guarantees that the dynamics is
preserved. However, the direct way can be unusable due to
$P_0$-growth. One can then increase $\beta$ and, simultaneously,
decrease the modulation depth $\mu$ as well as $\zeta$, $\gamma$
(e.g. by means of mode growth in a gas-filled resonator) in
accordance with the scaling rules but keeping $P_0$ below $P_{th}$.

\section{\label{s3}Brief comparison of the models}

Section \ref{s1} reports the development of a model of a CPO
mode-locked by an ideally saturable absorber. Such a model is
applicable to both the ANDi fiber oscillator \cite{wise2} and
solid-state CPO mode-locked by SESAM. Simultaneously, there are
models of a CPO based on the nonlinear cubic-quintic CGLE
\cite{kalash1,kalash3,wise1,akh1,akh2}. Such models take into account the
SAM saturation (i.e. the decrease of SAM with power overgrowth),
which is important in  high-power Kerr-lens mode-locked oscillators
as well as  fiber oscillators mode-locked by a polarization
modulator.

As a result of cubic-quintic SAM, a variety of the nonlinear regimes
appear \cite{akhmed}. In particular, the flat-top pulse envelope
develops. Since the cubic-quintic SAM confines the pulse peak power,
there is the limit for the pulse energy growth \cite{kalash1}. It
should be noted, that the SAM saturation parameter (i.e. parameter
defining the quintic term in the cubic-quintic SAM) is not truly a
free parameter in a real-world oscillator. It is closely related to
the inverse saturation power $\varsigma$ or to the self-focusing
power in a Kerr-lens mode-locked oscillator (see \cite{kalash1}). In
the last case, it is not possible to manipulate with SAM and SPM
independently because both processes result from the nonlinear
refraction in an active medium. Therefore such a type of SAM is out
of use for the microjoule oscillators.

One has to note, that underestimation of the SAM saturation
parameter in the cubic-quintic CGLE can lead to a huge CSP spectral
width and peak power (e.g. see \cite{akh2}). Constraint on such a
power growth can be provided by the gain saturation and, as a result
of $|a|$-growth, the CPO operates in the $-$ branch of CPO (in terms
of this article, see above). Switching to the $+$ branch of the CPO
can be provided by i) GDD decrease, ii) energy growth, or iii)
spectral filter band narrowing.

Finally, the low-energy ($-$ branch) sector (as well as the large
positive GDD sector) of CPO can be described by the nonlinear cubic
CGLE (i.e. by dissipative generalization of the nonlinear
Schr\"{o}dinger equation). Such a model is useful for a fiber
oscillator. For the cubic SAM, approach to the $+$ branch of CSP due
to the GDD decrease can cause the collapse-like instability because
there is no the peak power confinement due to SAM saturation.

An important issue is the  appearance of concave spectra like those
in Figs. \ref{fig4}, \emph{D} and \ref{fig6}, \emph{D}. It is known
that, for the SAM described by the nonlinear cubic-quintic CGLE, the
CSP has either parabolic- or finger-top spectra
\cite{kalash1,wise1}. However, both numerical simulations
\cite{wise1,akh2} and experiment \cite{wise1,wise2} demonstrate more
complicated spectral profiles: concave and convex-concave. In Ref.
\cite{wise1} the analytical concave spectra appear only for the
cubic-quintic SAM, which enhances the collapse-like instability
(i.e. there is no SAM saturation). In Ref. \cite{kalashn}, the
stable analytical CSPs with concave spectra are obtained for nonzero
quintic SPM. For the SAM type considered in the present work, the
analytical concave spectral profiles are not admissible because the
first equation of (\ref{sol2}) demonstrates that $\Delta'^2 \leq
3(1+a)$. Hence, the existence of concave spectra cannot be explained
on the basis of the present dissipative soliton models (with the
exception of the model presented in \cite{kalashn}).

Figures \ref{fig4}, \ref{fig6} show that the concavity grows with
spectrum broadening when $1/(2\Delta)^2$ tends to $\alpha$.
Simultaneously, there is no  concave spectrum, when SPM is
suppressed (Fig. \ref{fig8}, where $\gamma < \mu \varsigma$).

\begin{figure}
\includegraphics[width=9.5cm]{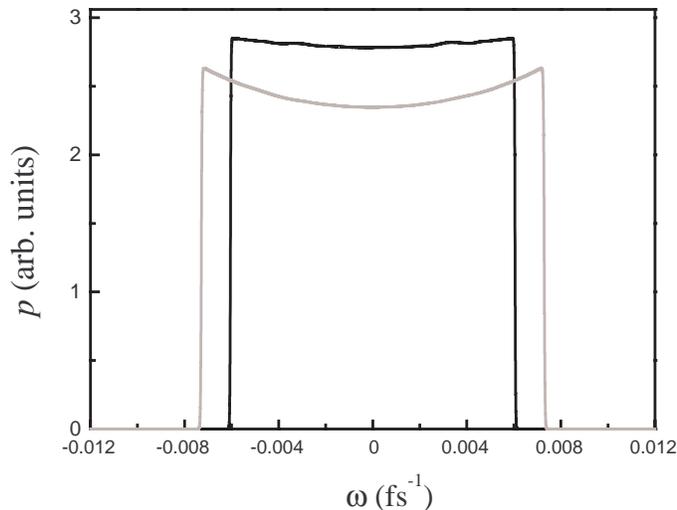}
\caption{\label{fig9} Numerical spectral profiles corresponding to
the model (\ref{GL}) (black curve) and its nonlinear cubic limit
(gray curve). Parameters correspond to the point $B$ in Fig.
\ref{fig3}.}
\end{figure}

Numerical analysis demonstrates that the appearance of concave
spectrum is not rooted in the character of SAM. Surprisingly,
however,  the concave spectrum profile exists even in the case of
the nonlinear cubic CGLE and develops with the spectrum broadening.
Figure \ref{fig9} shows the spectra corresponding to point \emph{B}
in Fig. \ref{fig3}, but for the different SAM types: $ {{\mu
\varsigma \left| A \right|^2 } \mathord{\left/
 {\vphantom {{\mu \varsigma \left| A \right|^2 } {\left( {1 + \varsigma \left| A \right|^2 } \right)}}} \right.
 \kern-\nulldelimiterspace} {\left( {1 + \varsigma \left| A \right|^2 } \right)}}
$ (black curve, see Eq. (\ref{GL})) 
and $ \mu \varsigma \left| A \right|^2 $ (nonlinear cubic model,
gray curve). It can be seen that the absence of SAM saturation in
the second case leads to spectrum broadening due to power growth
($\Delta^2 \propto P(0)$, see Eqs. (\ref{sol2},\ref{cubic})).
Analytical estimation for $\Delta$ from Eq. (\ref{cubic}) gives
0.0076 fs$^{-1}$ vs. numerical value 0.0074 fs$^{-1}$. As a result
of the spectrum broadening, the concavity increases. The
$P(t)$-profile does not have any anomalies. Thus, growth of
spectral components at the spectrum edges can be treated as a
fundamental feature of the regime considered (see also Ref.
\cite{kalash4}), without obligatory addressing of
higher-order dispersions \cite{oe}.

\section{Conclusions} \label{conclusion}

Analytical theory of the CSP has been developed. The theory treats
the CSP as a solitary pulse solution of the generalized nonlinear
CGLE. The main advantage of the theory developed is the possibility
of representing the CPO parametrical space in the form of a
two-dimensional master diagram. As a result, the CSP characteristics
become easily traceable. It has been demonstrated that both ANDi
fiber and chirped-pulse solid-state oscillators can be described
from a uniform standpoint and represented on a unified master
diagram. The main difference between them is that they realize
mainly two different branches of the CSP solution. Such branches
differ in the energy and dispersion scaling rules as well as in the
behavior of the CSP parameters.

Comparison with the results of numerical simulations has shown that
the analytical solution provides a good approximation of the
spectrum shape, which is truncated and has a flat or parabolic top.
The approximation is quite precise even in the case  where the
net-gainband is so narrow that the squared inverse bandwidth verges
towards the GDD. This provides an adequate description of both ANDi
fiber and thin-disk narrowband solid-state chirped-pulse
oscillators. Thus, the theory allows the CPO
characteristics to be optimized and demonstrates the feasibility of  at least,
sub-mJ energy scaling in a thin-disk CPO.

\begin{acknowledgments}

This work was supported in part by Deutsche Forschungsgemeinschaft
through the DFG cluster of excellence Munich Centre for Advanced
Photonics (www.munich-photonics.de). Author V.L.K. acknowledges
support from the Max Planck Society (Germany) and Austrian Fonds zur
F\"{o}rderung der wissenschaftlichen Forschung (project P20293).
Authors acknowledge A.Fernandez for the technical support.
\end{acknowledgments}

\end{document}